\documentclass[12pt,preprint]{aastex}

\begin{document}

\title{Multiple Wavelength Variability and Quasi-Periodic Oscillation of PMN J0948+0022}
\author{Jin Zhang\altaffilmark{1,2}, Hai-Ming Zhang\altaffilmark{2}, Yong-Kai Zhu\altaffilmark{2}, Ting-Feng Yi \altaffilmark{3}, Su Yao \altaffilmark{4}, Rui-Jing Lu\altaffilmark{2}, En-Wei Liang\altaffilmark{2}}
\altaffiltext{1}{Key Laboratory of Space Astronomy and Technology, National Astronomical Observatories, Chinese Academy of Sciences, Beijing 100012, China; jinzhang@bao.ac.cn}
\altaffiltext{2}{Guangxi Key Laboratory for Relativistic Astrophysics, Department of Physics, Guangxi University, Nanning 530004, China}
\altaffiltext{3}{Department of Physics, Yunnan Normal University, Kunming 650500, China}
\altaffiltext{4}{Kavli Institute for Astronomy and Astrophysics, Peking University, Beijing 100871, China; KIAA-CAS Fellow }

\begin{abstract}
We present a comprehensive analysis of multiple wavelength observational data of the first GeV-selected narrow-line Seyfert 1 galaxy PMN J0948+0022. We derive its lightcurves in the $\gamma$-ray and X-ray bands from the data observed with \emph{Fermi}/LAT and \emph{Swift}/XRT, and make the optical and radio lightcurves by collecting the data from the literature. These lightcurves show significant flux variations. With the LAT data we show that this source is analogue to typical flat spectrum radio quasars in the $L_\gamma-\Gamma_\gamma$ plane, where $L_\gamma$ and $\Gamma_\gamma$ are the luminosity and spectral index in the LAT energy band. The $\gamma$-ray flux is correlated with the V-band flux with a lag of $\sim 44$ days, and a moderate quasi-periodic oscillation (QPO) with a periodicity of $\sim 490$ days observed in the LAT lightcurve. A similar QPO signature is also found in the V-band lightcurve. The $\gamma$-ray flux is not correlated with the radio flux in 15 GHz, and no similar QPO signature is found in a confidence level of 95\%. Possible mechanisms of the QPO are discussed. We propose that gravitational wave observations in the future may clarify the current plausible models for the QPO.
\end{abstract}


\section{Introduction}           
PMN J0948+0022 (redshift $z=0.584$; Zhou et al. 2003) is the first $\gamma$-ray source detected with \emph{Fermi}/LAT (Large Area Telescope; Abdo et al. 2009a) among narrow-line Seyfert 1 galaxies (NLS1s). Seven NLS1s have been detected with \emph{Fermi}/LAT so far (Abdo et al. 2009b; D'Ammando et al. 2012; D'Ammando et al. 2015; Yao et al. 2015), and PMN J0948+0022 is still the brightest one in the GeV band among these GeV-selected NLS1s. Both the radio power and radio loudness (RL) of this source are analogous to the classical radio quasars, i.e., $P_{\rm rad} > 10^{26}$ W Hz$^{-1}$ and $R_{\rm RL} > 1000$ (Zhou et al. 2003). It displays blazar characteristics and may also host a relativistic jet (Zhou et al. 2003; Yuan et al. 2008). This was confirmed with the \emph{Fermi}/LAT observations. Its broadband spectral energy distributions (SEDs) observed in different campaigns can be well explained by the one-zone leptonic models (Abdo et al. 2009a; Foschini et al. 2012; Zhang et al. 2013), and the derived jet properties are consistent with flat spectrum radio quasars (FSRQs, e.g., Sun et al. 2015).

Significant flux variations were observed with \emph{Fermi}/LAT in PMN J0948+0022 (e.g., Foschini et al. 2012; Sun et al. 2014, 2015), which is similar to the characteristics observed in blazars. It is also interesting that the flux variations of some blazars also show the phenomenon of quasi-periodic oscillation (QPO), such as the 12-year QPO in OJ 287 (e.g., Sillanpaa et al. 1988, 1996), the $1884\pm88$-day QPO in PG 1302--102 (Graham et al. 2015), the 630$\sim$640-day QPO in PKS 2155-304 (Sandrinelli et al. 2014; Zhang et al. 2017a), the $2.18\pm0.08$-year QPO in PG 1553+113 (Ackermann et al. 2015), etc. By studying the variability of six blazars in optical-near-infrared and $\gamma$-ray bands, Sandrinelli et al. (2016) also suggested that a year-like QPO may be often observed in blazars.

This paper presents a comprehensive analysis of multiple wavelength observational data of PMN J0948+0022. It is known that the flux variation of blazars is usually accompanied by the variation of the spectral index (e.g., Cui 2004; Massaro et al. 2008; Tramacere et al. 2009; Nalewajko 2013). We focus on revealing its flux variations in multiple wavelengthes and spectral variation features. We try to find out the possible QPO signature from the long-term observation data in the $\gamma$-ray, optical and radio bands. As described in Section 2, we derive its lightcurves in the $\gamma$-ray and X-ray bands from the data observed with \emph{Fermi}/LAT and \emph{Swift}/XRT, and make the optical and radio lightcurves by collecting the data from the literature. The cross-correlation analysis of variability among multiple wavelength lightcurves is given in Section 3. With the well-sampled observation data of \emph{Fermi}/LAT, we investigate the flux variation and spectral evolution in the GeV band in Section 4. Searching for the possible QPO signature in multiple wavelength lightcurves is presented in Section 5. A discussion is given in Section 6 and conclusions are reported in Section 7.

\section{Data Reduction}

\subsection{\emph{Fermi}/LAT Data Analysis}

\begin{figure*}
\includegraphics[angle=0,scale=0.45]{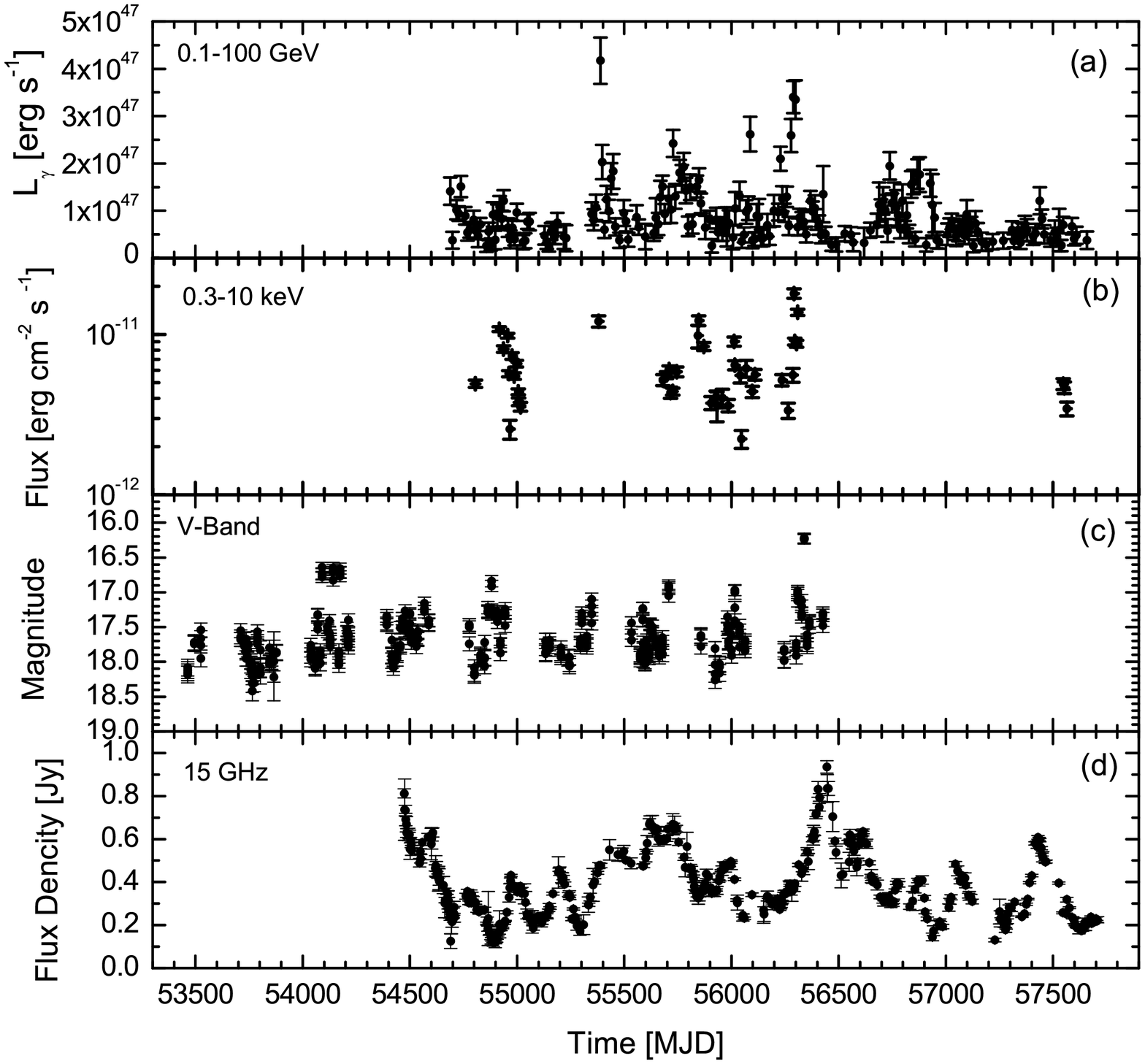}\\

\caption{Lightcurves in multiwavelengths, from top: \emph{Fermi}/LAT $\gamma$-ray, \emph{Swift}/XRT X-ray, optical V-Band, and radio 15 GHz. The LAT lightcurve with TS$>$5 and timebin=10-day are from 2008 August 5 (MJD 54683) to 2016 December 31 (MJD 57753). The optical data are taken from the CRTS. The radio data are from the radio monitoring program with the 40 m telescope at the OVRO. }\label{LC_all}
\end{figure*}

The Pass 8 \emph{Fermi}/LAT data of PMN J0948+0022 with a temporal coverage from 2008 August 5 (Modified Julian Day, MJD 54683) to 2016 December 31 (MJD 57753) are downloaded from the Fermi Science Support Center (\emph{http://fermi.gsfc.nasa.gov/ssc}). The data analysis is performed using the standard ScienceTools v10r0p5 software package. Events from 100 MeV to 100 GeV within a circular region of interest of $10^\circ$ radius centered on the location of PMN J0948$+$0022 (RA=147.238837, Dec=0.373767) are selected with P8R2\_SOURCE\_V6 instrument response functions. We use the standard unbinned maximum likelihood fit technique and a power-law spectral model to analyze each time bin. The integrated flux and photon spectral index of the power-law are taken as free parameters during the fitting. In order to eliminate the contamination from the $\gamma$-ray-bright Earth limb, the events with zenith angle $>$100$^\circ$ are excluded. The recently released files gll\_iem\_v06 and iso\_P8R2\_SOURCE\_V6\_v06 are used to model the Galactic and isotropic diffuse emission. The significance of the $\gamma$-ray signal from the source is evaluated with the maximum-likelihood test statistic (TS). The LAT lightcurve with TS$>$5 in time-bins of 10-day is shown in panel (a) of Figure \ref{LC_all}.

\subsection{\emph{Swift}/XRT Data Analysis}

PMN J0948+0022 was observed on 2008 December 5 with the X-Ray Telescope (XRT) on board the {\em Swift} satellite, soon after the detection of the $\gamma$-ray emission from this source (Abdo et al. 2009a). During the multi-wavelength monitoring campaigns in 2009, there were 11 observation snapshots (Abdo et al. 2009c). An exposure of 1615 seconds was carried on 2010 July 3 (Foschini et al. 2012). Ten more exposures were performed in 2011, as part of a monitoring program linked to the Effelsberg radio observations(Foschini et al. 2012). There were fifteen observation snapshots in 2012, and three exposures were performed in 2013 and in 2016, respectively. In total, there are 44 observation snapshots of \emph{Swift}/XRT with each exposure time of 1--5 ks from 2008 to 2016. We download the XRT data from \emph{http://www.swift.ac.uk/archive/}. We use the \emph{Swift} software (HEASoft v.6.17 package and the CALDB version updated on 2016 June 9) to deduce XRT data of PMN J0948+0022. The spectra in the XRT band (0.3--10 keV) are fitted  with a single power-law model and the Galactic absorption corresponding to a hydrogen column density is fixed at $N_{\rm H}=5.22\times10^{22}$ cm$^{-2}$ (Kalberla et al. 2005). The derived XRT lightcurve is shown in the panel (b) of Figure \ref{LC_all}.

\subsection{Radio and Optical Data}
We collect the V-band and 15 GHz radio data from the literatures and show them in the panels (c) and (d) of Figure \ref{LC_all}. The V-band data, which  covers $\sim$3000 days (from 2005 April 4 to 2013 May 14; MJD 53464--56426), are taken from the Catalina Real-Time Transient Survey (CRTS; \emph{http://crts.caltech.edu/}; Drake et al. 2009; Mahabal et al. 2011), which is an unfiltered optical survey for transients. The 15 GHz data are obtained from the continual observations with the 40 m Owens Valley Radio Observatory (OVRO, \emph{http://www.astro.caltech.edu/ovroblazars/data/data.php}) radio telescope. OVRO instrumentation, data calibration, and reduction are described in Richards et al. (2011). The OVRO supports an ongoing blazar monitoring program of \emph{Fermi} satellite, hence the lightcurve in 15 GHz almost covers the same observation period of \emph{Fermi}/LAT.

\section{Cross Correlations among Multiple Wavelength Lightcurves}

\begin{figure*}
\includegraphics[angle=0,scale=0.37]{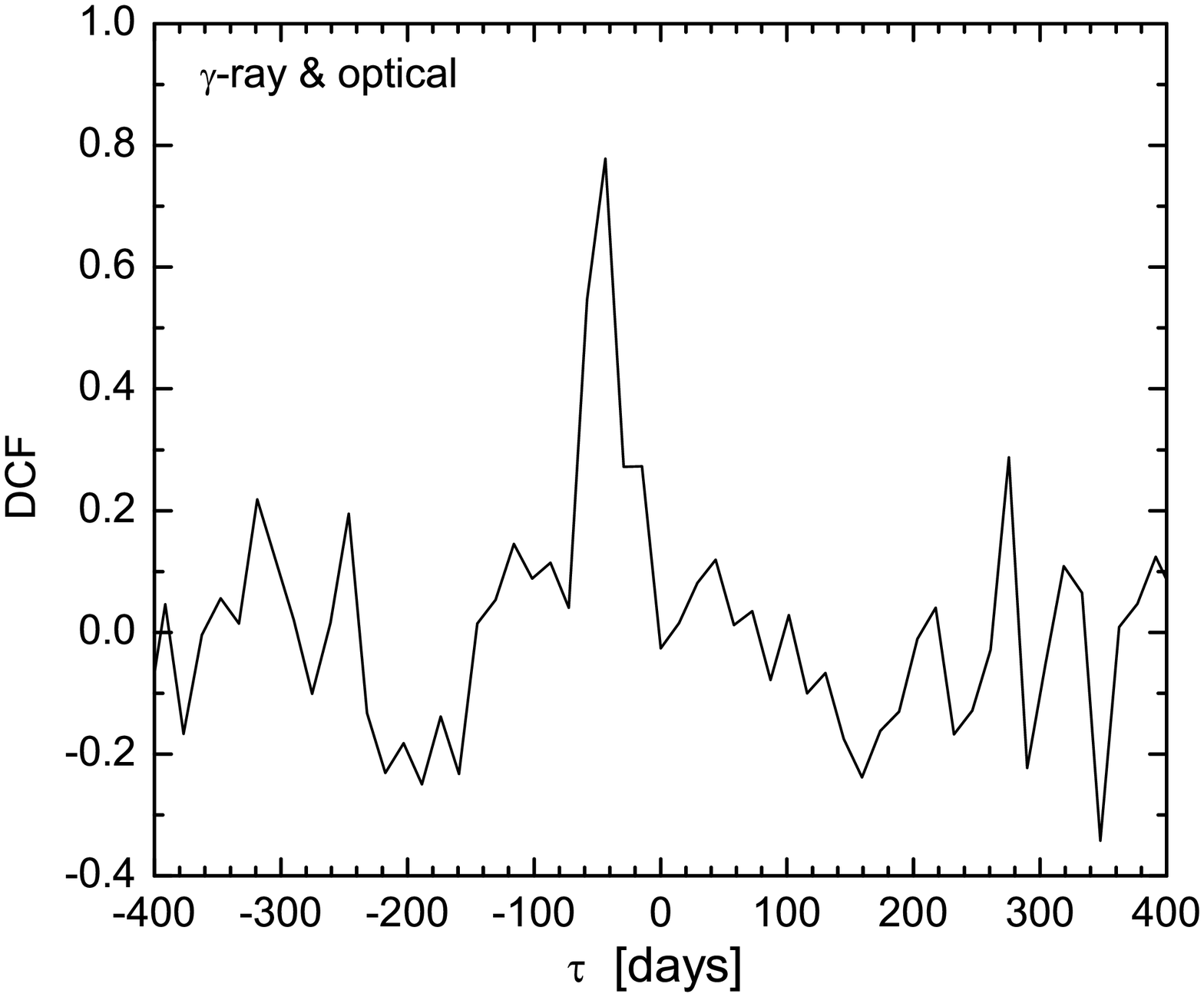}
\includegraphics[angle=0,scale=0.37]{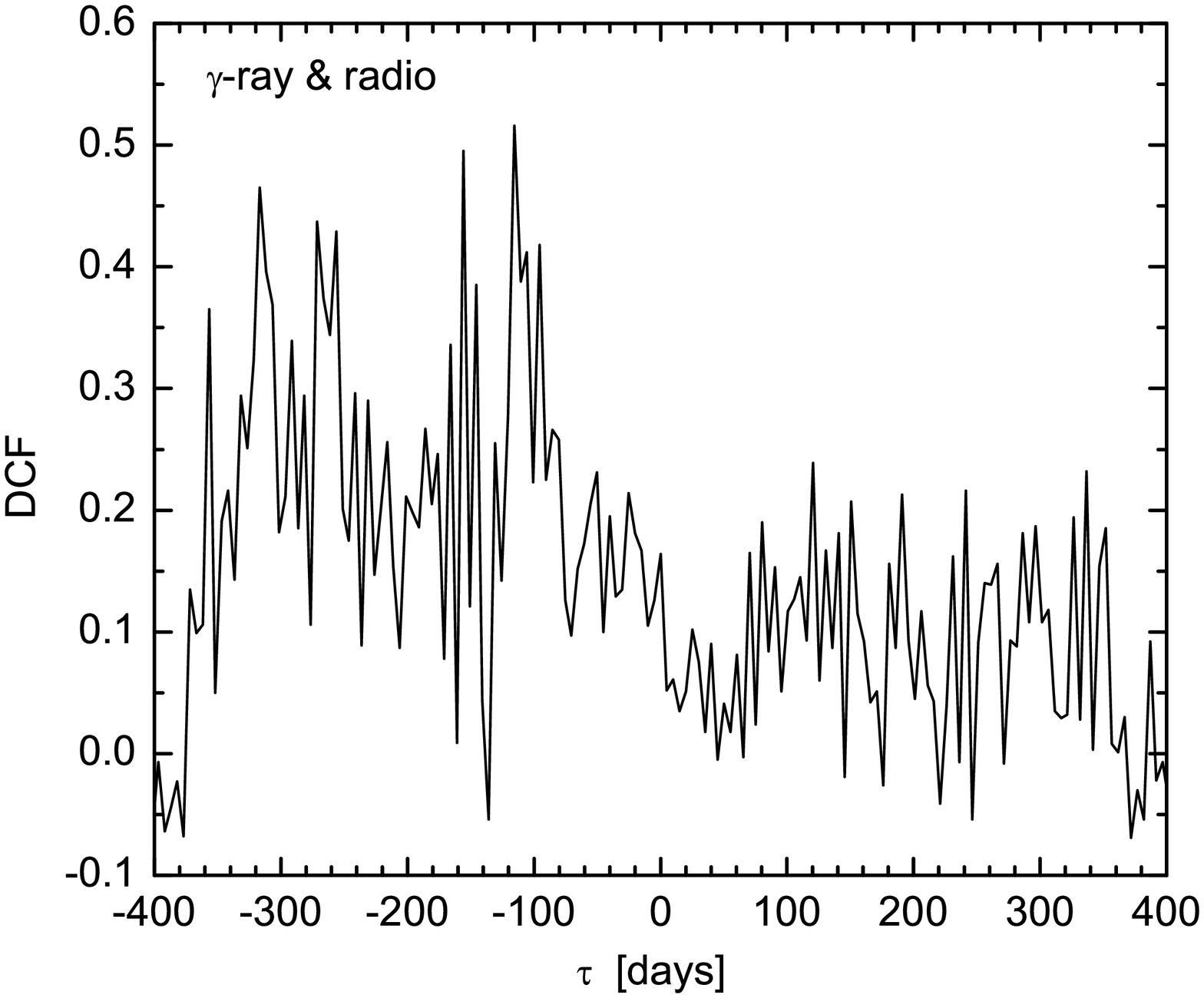}\\
\caption{DCF results between $\gamma$-ray (time-bin=10-day) with optical V-band and radio 15 GHz, where the observation data in the same temporal coverage (from MJD 54776 to MJD 56427) are used to do DCF analysis.}\label{DCF}
\end{figure*}

As shown in Figure \ref{LC_all}, PMN J0948+0022 shows significant flux variation in multiple wavelength. It is very active without showing any quiescent stage during the data coverage from MJD 53500 to MJD 57500. We first analyze the cross correlations of the $\gamma$-ray flux to optical and radio fluxes with the data observed in the same temporal coverage. Being due to the poorly sampling, we do not make correlation analysis for the X-ray data. Our analysis results with discrete cross-correlation function (DCF, Edelson \& Krolik 1988) are presented in Figure \ref{DCF}. One can observe that the $\gamma$-ray flux is correlated with the optical flux with a lag of $\sim 44 $ days, and the correlation coefficient is 0.78. The correlation of variability between $\gamma$-ray and optical bands for blazars has been widely reported (e.g., Arlen et al. 2013; Jorstad et al. 2013; Ghisellini \& Tavecchio 2009).

The DCF for the
$\gamma$-ray and radio data shows many peaks, but no statistical correlation can be claimed for the $\gamma$-ray and radio data. It is possible that the radio emission is radiated by an electron population different from that for the $\gamma$-ray and optical emission since the radio emission at 15 GHz of the $\gamma$-ray emitting region would not be detected, being due to the synchrotron-self-absorption effect.

\section{Flux and Spectral Variations in the GeV Energy Band}

\begin{figure*}
\includegraphics[angle=0,scale=0.45]{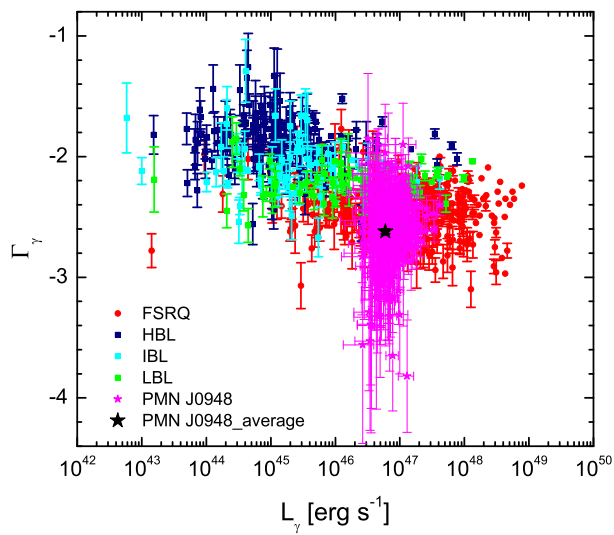}\\

\caption{$\Gamma_{\gamma}$ as a function of $L_{\gamma}$, where the \emph{Fermi}/LAT observation data of PMN J0948+0022 with timebin=10-day are presented and the average of the 8-year \emph{Fermi}/LAT observation data for PMN J0948+0022 is marked as a \emph{black star}. The data of blazars, which are the average values of the first 4 yr of science data from the \emph{Fermi}/LAT, are taken from Ackermann et al. (2015, see their Figure 14), where HBL, IBL, and LBL indicate high-frequency-peaked BL Lac, intermediate-frequency-peaked BL Lac, and low-frequency-peaked BL Lac, respectively.  }\label{L-G}
\end{figure*}

The flux variation is generally accompanied by the spectral index variation for blazars. In this section we analyze the correlation between the $\gamma$-ray luminosity ($L_\gamma$) and the photon spectral index ($\Gamma_\gamma$) using the observation data of \emph{Fermi}/LAT for PMN J0948+0022, where $L_\gamma$ and $\Gamma_\gamma$ are derived with the time-bins of 10-days. Figure \ref{L-G} illustrates the variation of PMN J0948+0022 in the $L_\gamma-\Gamma_\gamma$ plane. In order to compare PMN J0948+0022 with blazars, we also show the {\em Fermi} blazars in the $L_\gamma-\Gamma_\gamma$ plane. The blazar data, which are the average values of the first 4 yr of science data from the \emph{Fermi}/LAT, are taken from Ackermann et al. (2015, see also their Figure 14). These blazars belong to the clean sample with confirmed redshift, including 414 FSRQs, 162 high-frequency-peaked BL Lacs, 69 intermediate-frequency-peaked BL Lacs, and 68 low-frequency-peaked BL Lacs. The $\gamma$-ray luminosity of PMN J0948+0022 varies from 2.5$\times10^{46}$ erg s$^{-1}$ to 4.2$\times10^{47}$ erg s$^{-1}$, and $\Gamma_\gamma$ varies from  $-4.26\pm1.03$ to $-1.85\pm0.54$ with a mean of -2.65 (see also Paliya et al. 2015). Our results illustrate that the radiation properties of this source in the GeV band are analogous to FSRQs (e.g., Sun et al. 2015).

\begin{figure*}
\includegraphics[angle=0,scale=0.36]{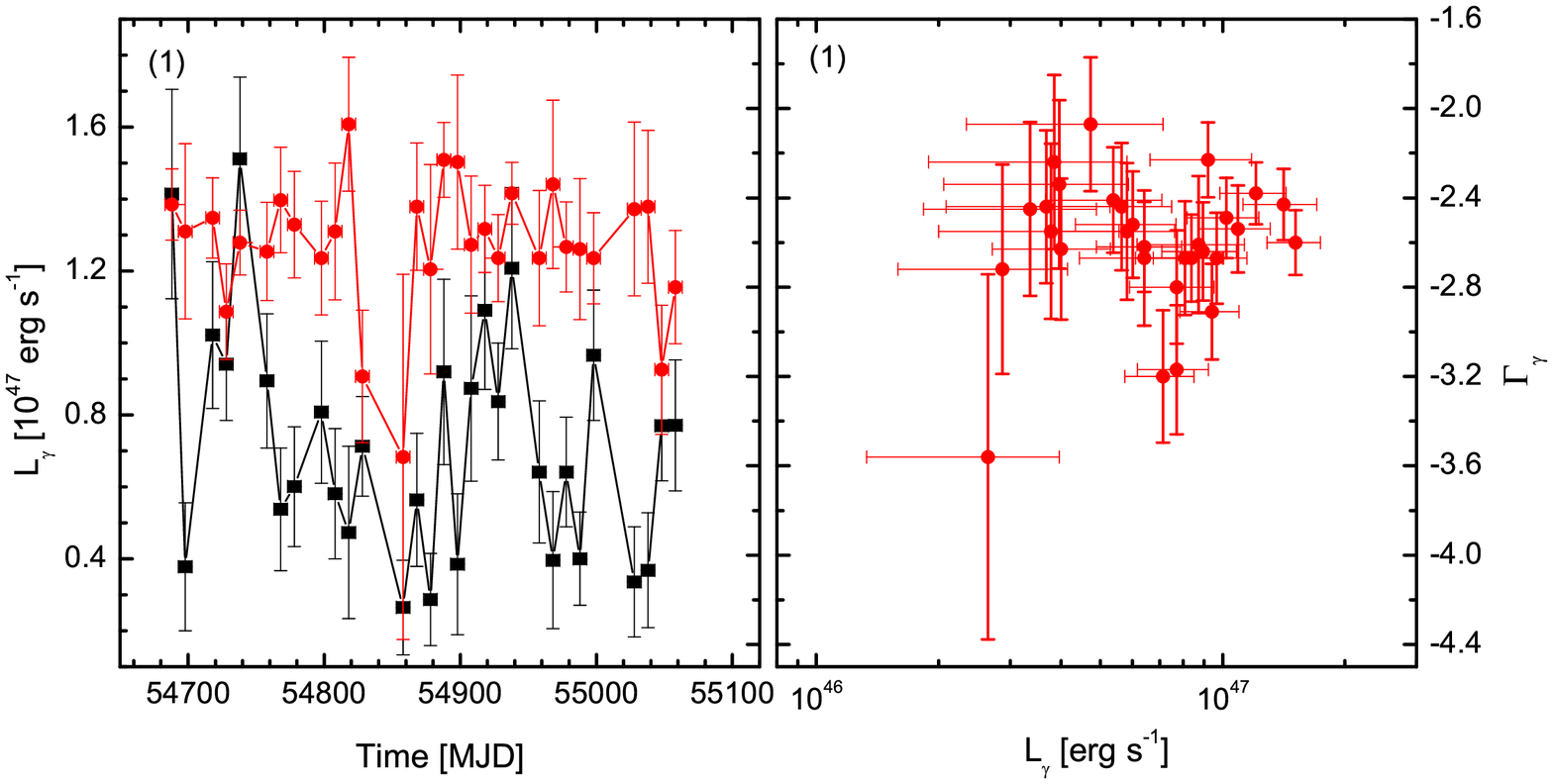}
\includegraphics[angle=0,scale=0.36]{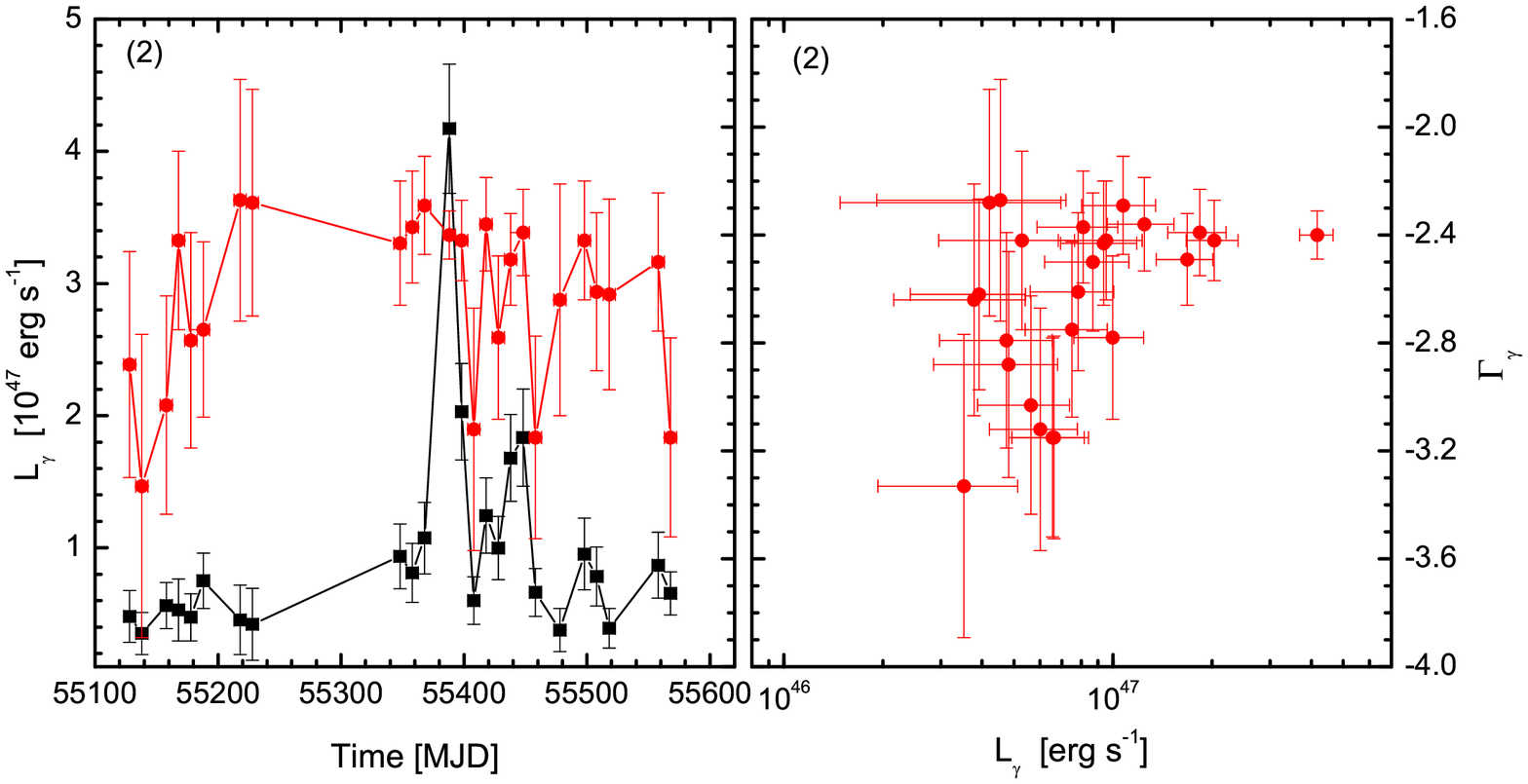}
\includegraphics[angle=0,scale=0.36]{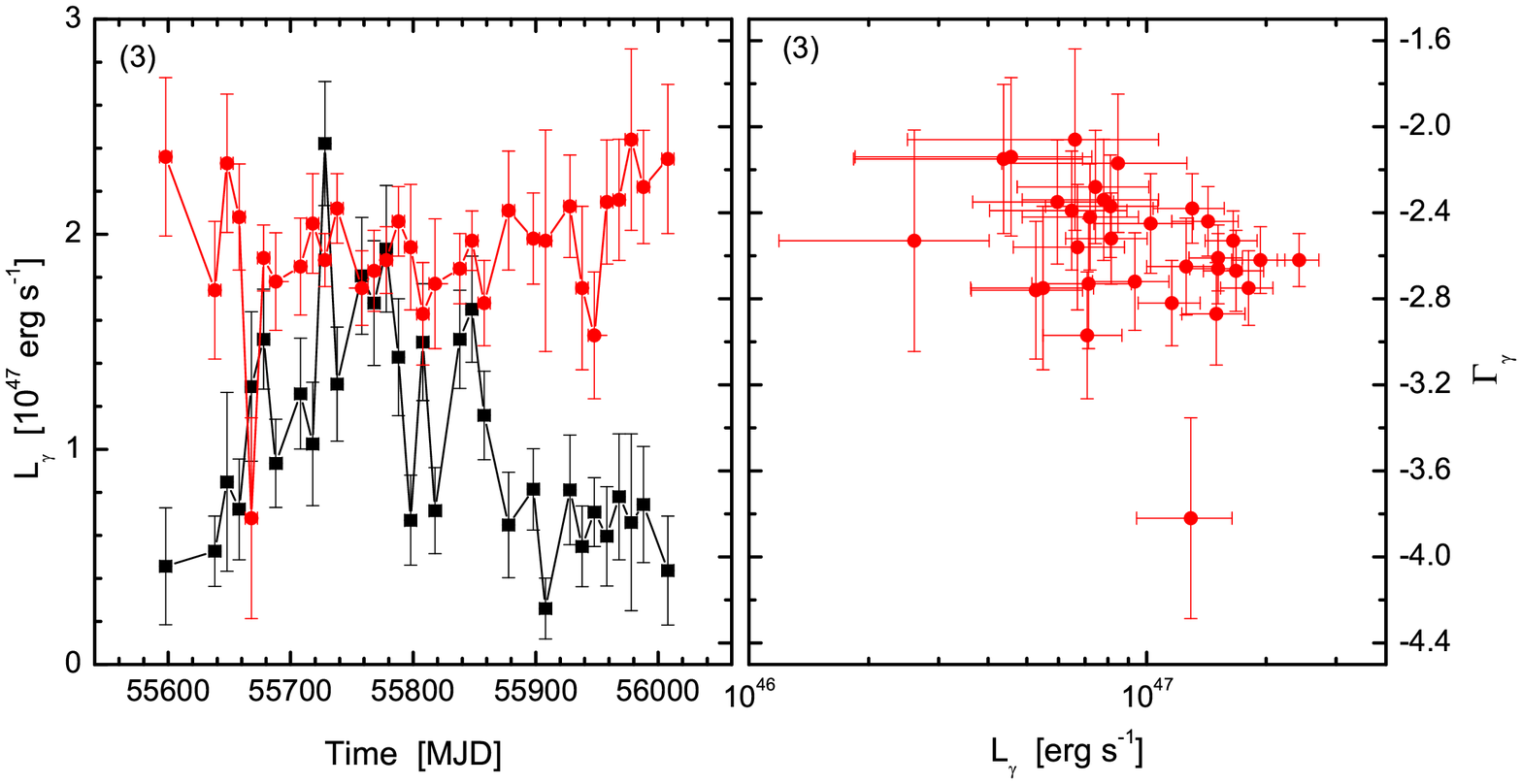}
\includegraphics[angle=0,scale=0.36]{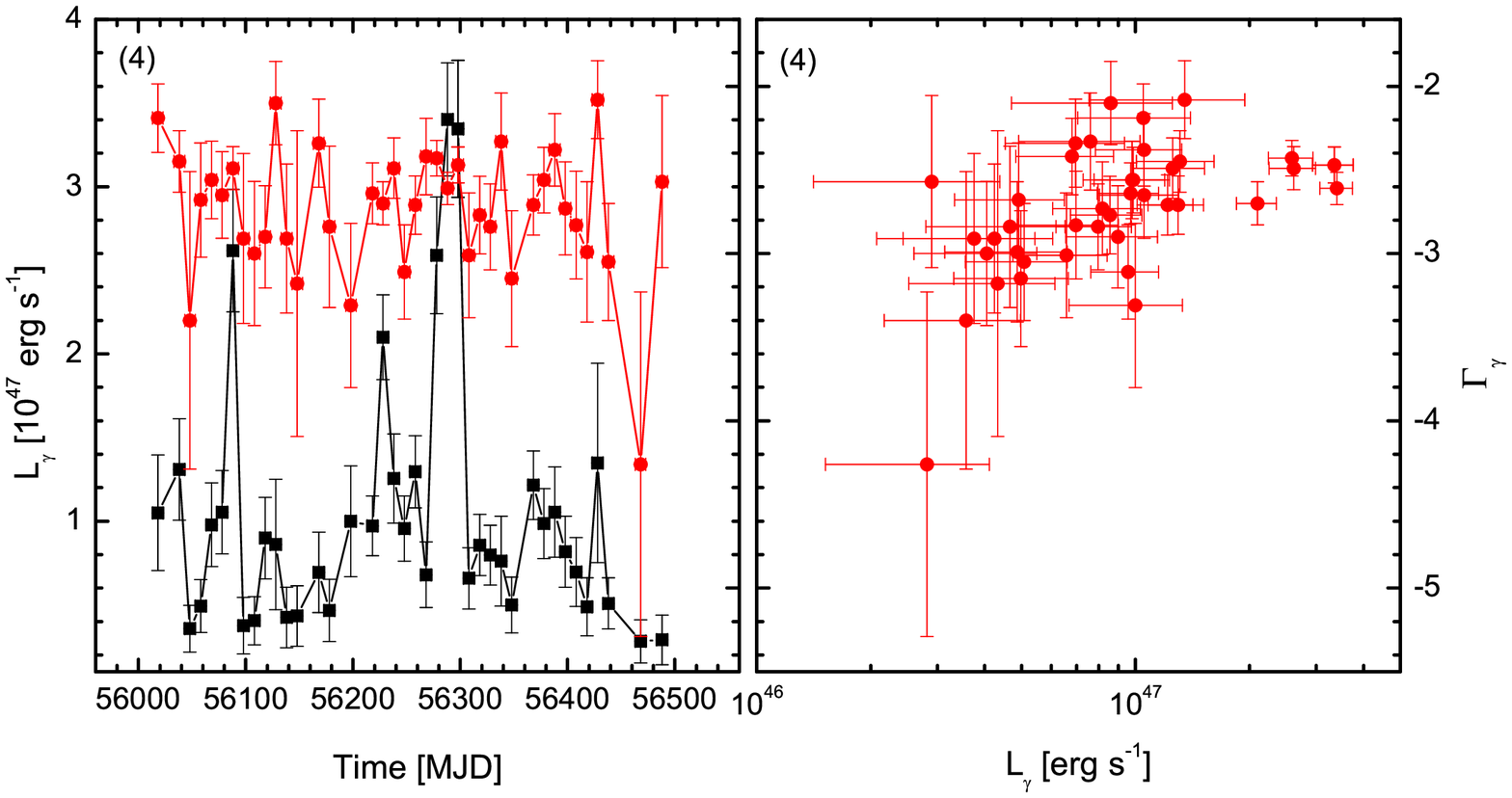}
\includegraphics[angle=0,scale=0.36]{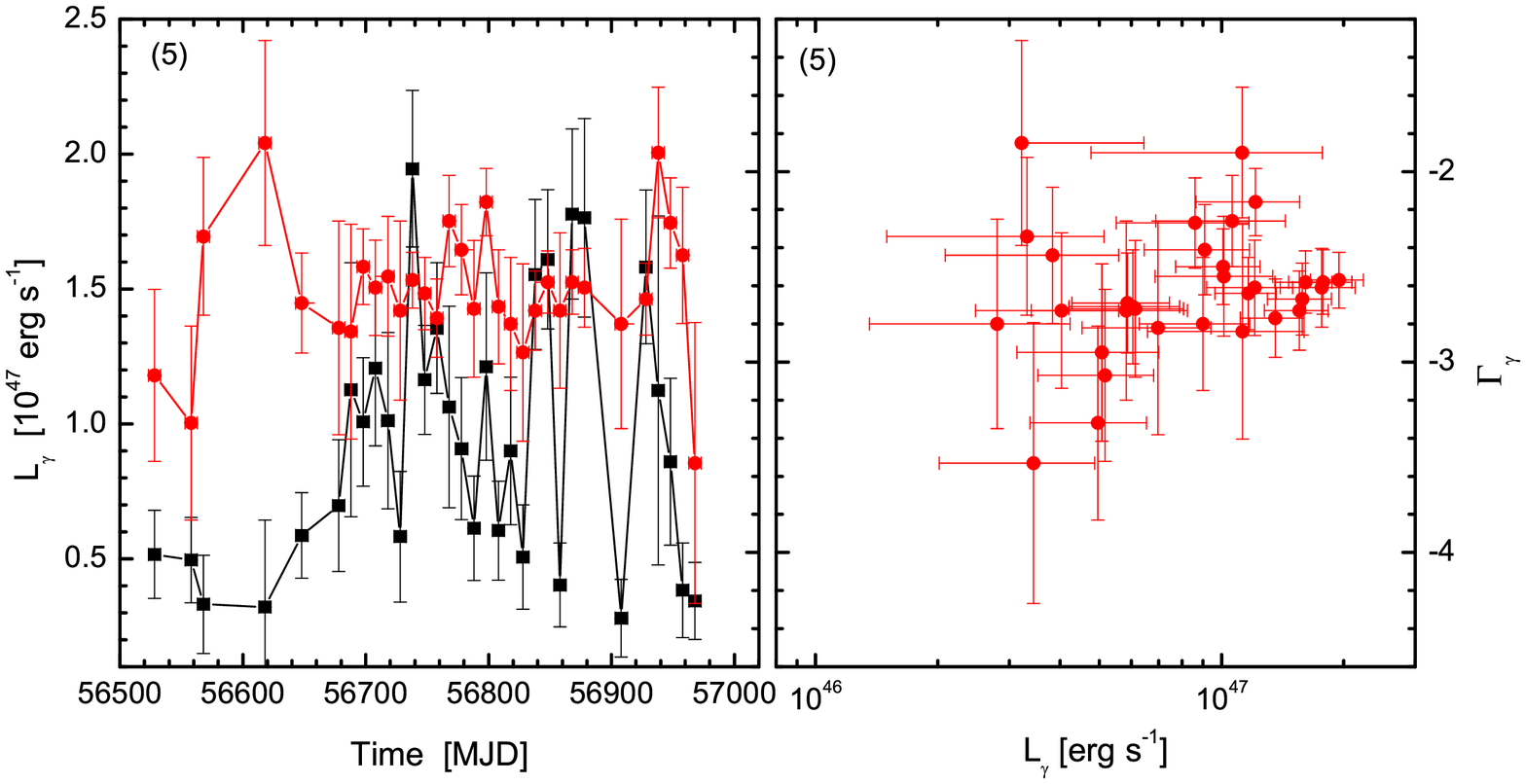}
\includegraphics[angle=0,scale=0.36]{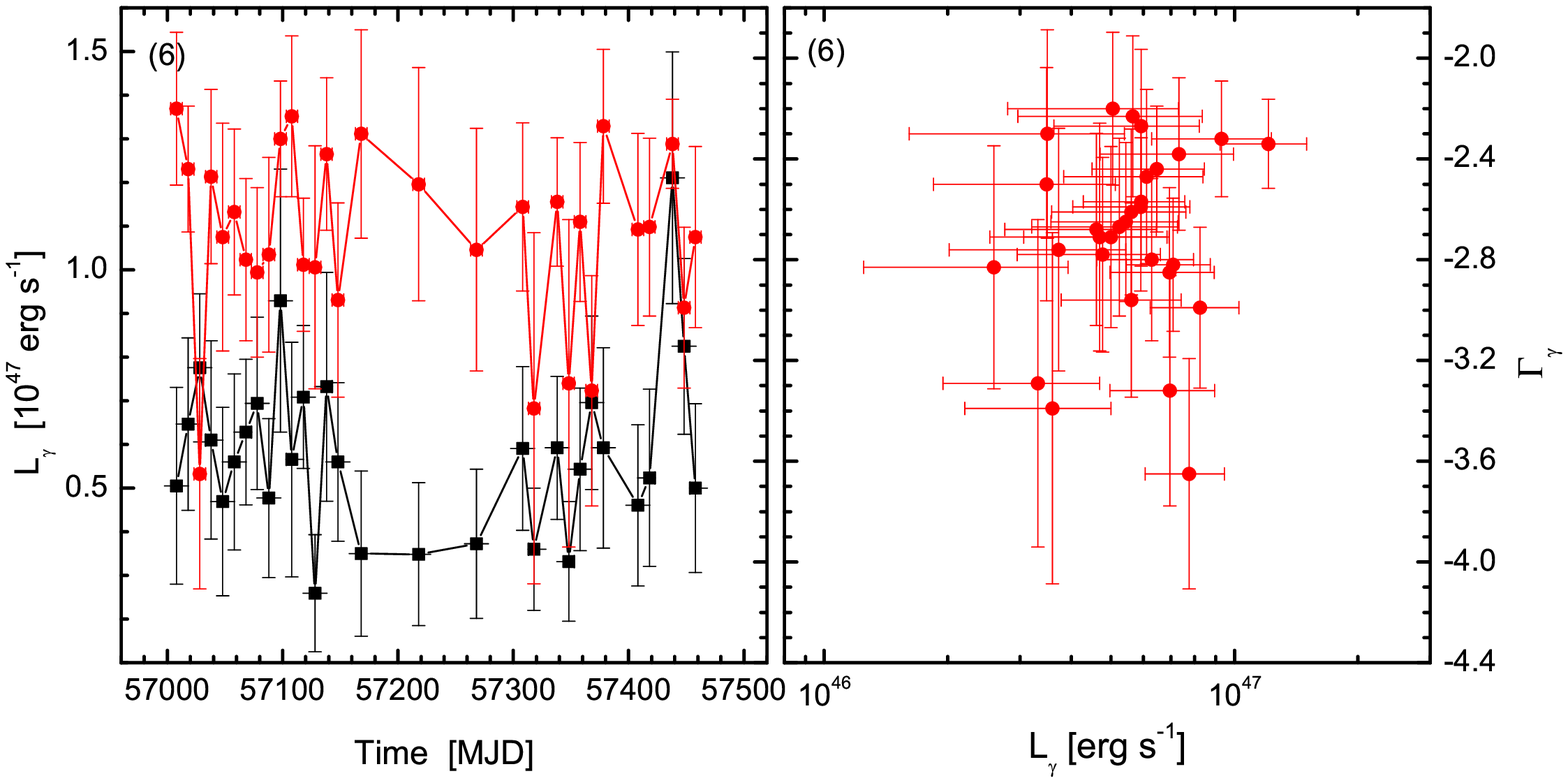}
\includegraphics[angle=0,scale=0.36]{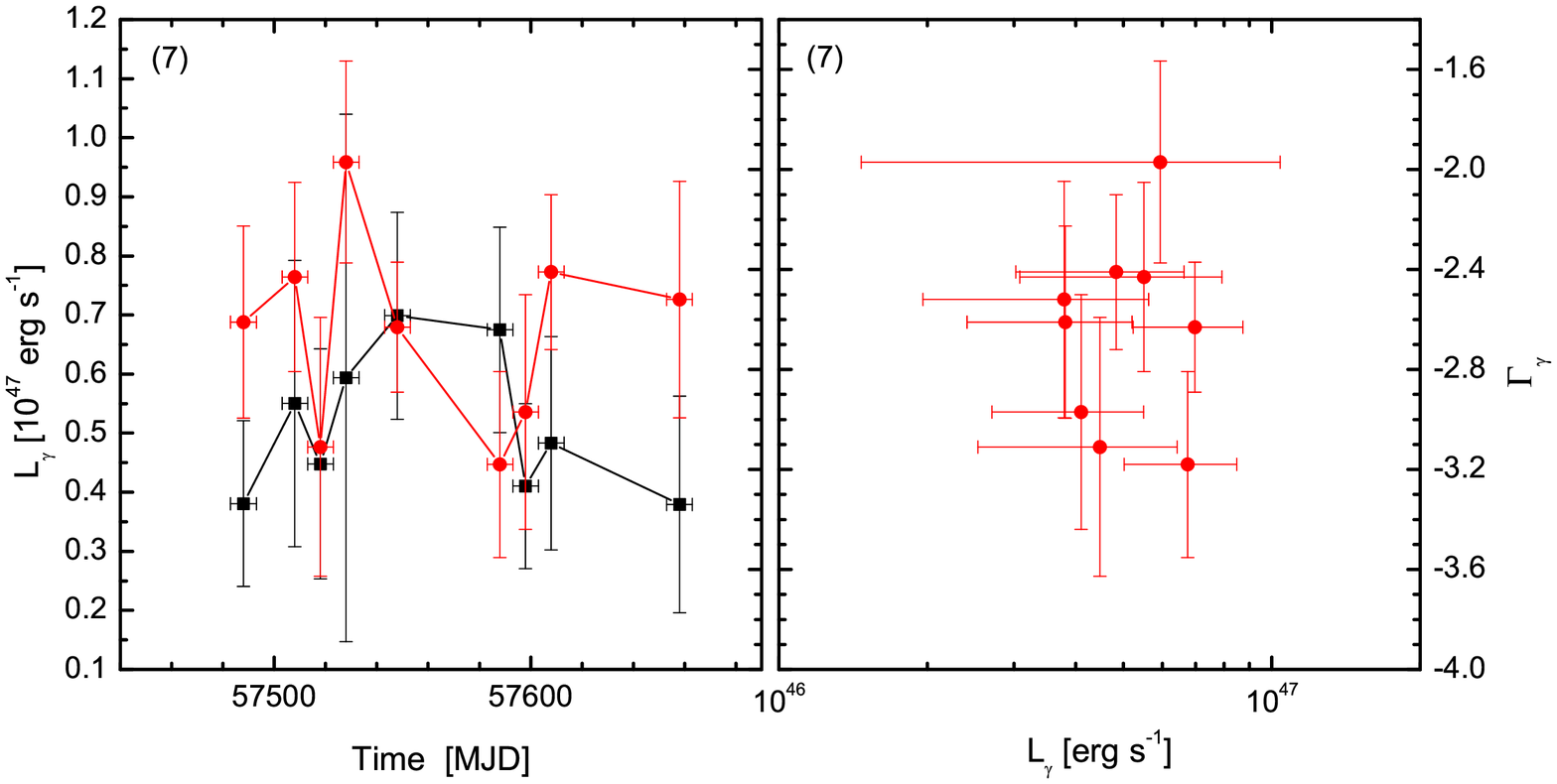}\\
\caption{The temporal variations of luminosity ($L_{\gamma}$, \emph{black squares}) and photon spectral index ($\Gamma_{\gamma}$, \emph{red circles}) with time-bins of 10-day, as well as $L_{\gamma}$ vs. $\Gamma_{\gamma}$ for the seven episodes.}\label{episodes}
\end{figure*}

\begin{figure*}
\includegraphics[angle=0,scale=0.36]{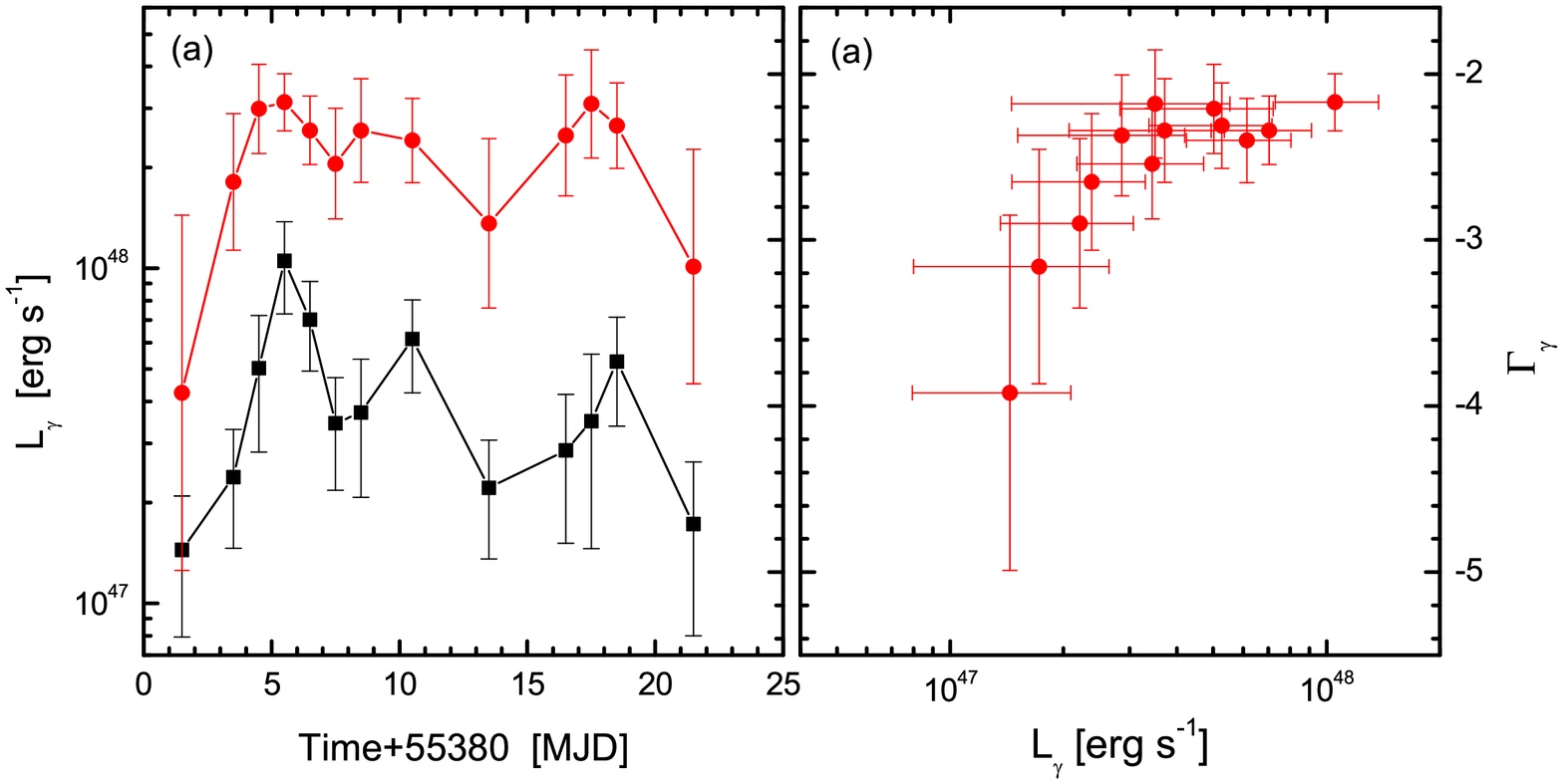}
\includegraphics[angle=0,scale=0.36]{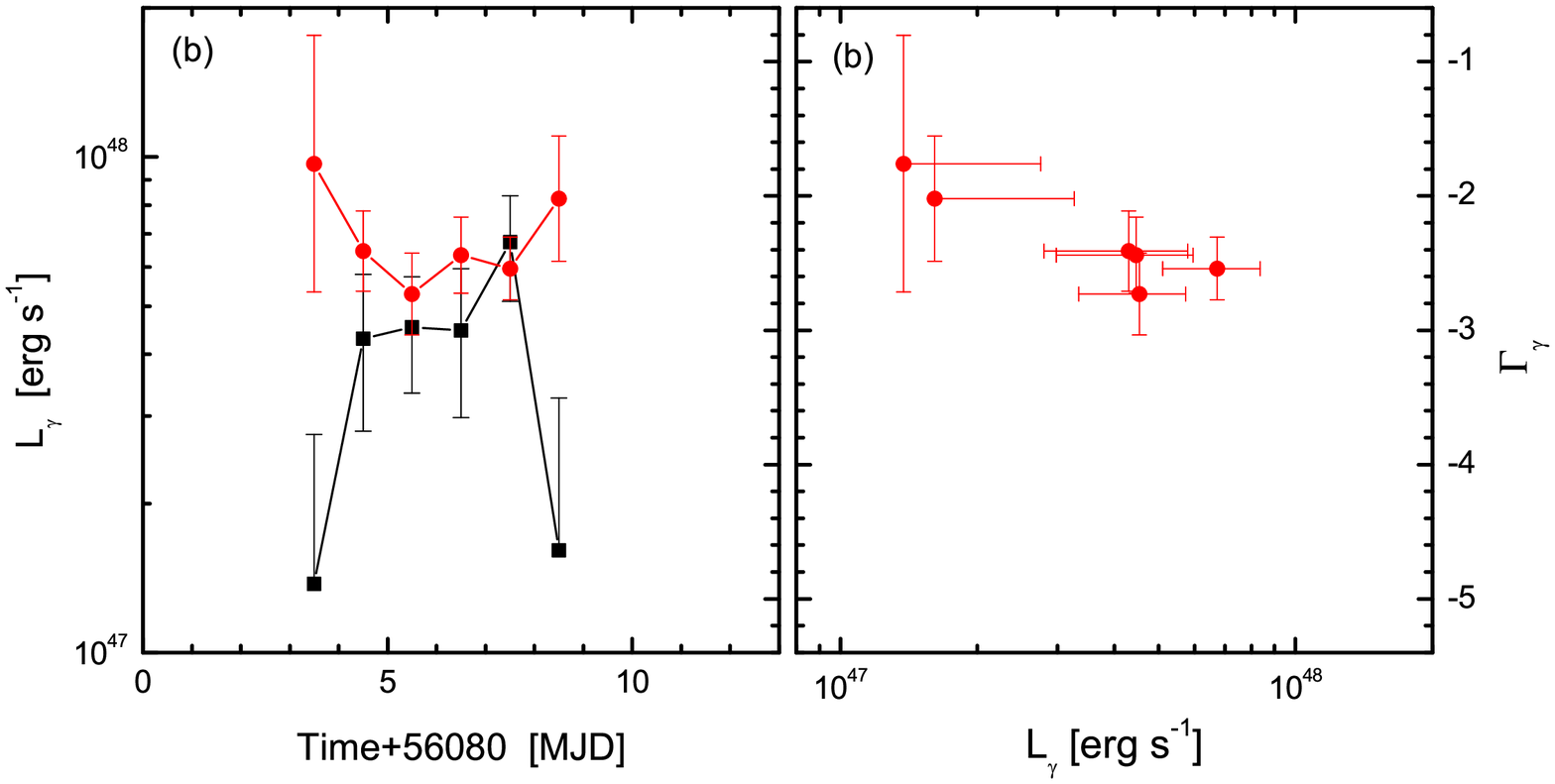}\\
\includegraphics[angle=0,scale=0.36]{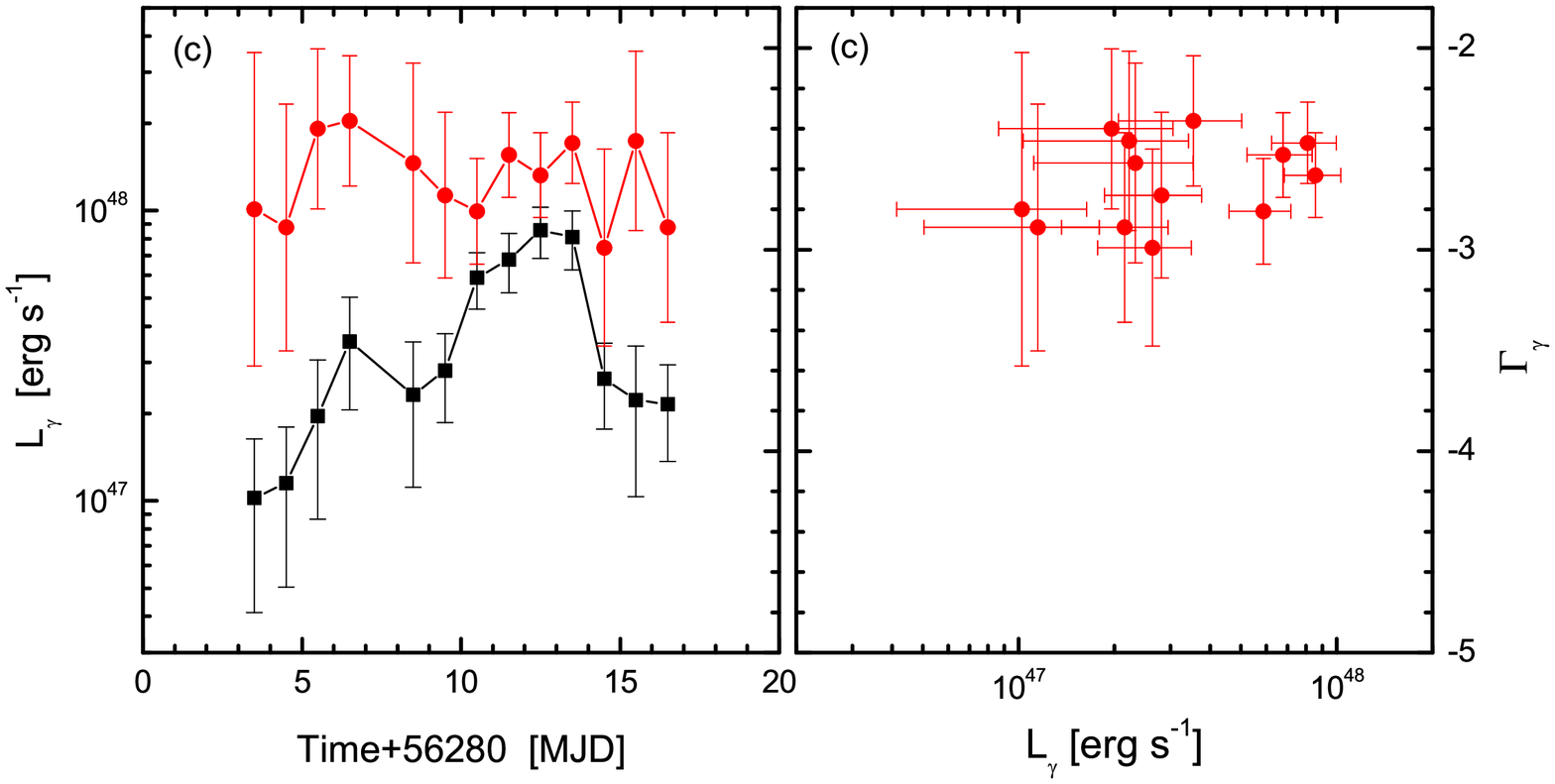}\\
\caption{The temporal variations of luminosity ($L_{\gamma}$, \emph{black squares}) and photon spectral index ($\Gamma_{\gamma}$, \emph{red circles}) with time-bins of 1-day, as well as $L_{\gamma}$ vs. $\Gamma_{\gamma}$ for the three flares (MJD [55381,55402], MJD [56083,56089], and MJD [56283,56297]).  }\label{flares}
\end{figure*}

We analyze the correlation between $L_\gamma$ and $\Gamma_\gamma$, but do not find any statistical correlation for the global LAT data with the Pearson correlation analysis method. By dividing the global LAT lightcurve into seven episodes, we show the temporal evolution of $\Gamma_\gamma$ and $\Gamma_\gamma$ as a function of $L_\gamma$ in Figure \ref{episodes}. Except for the episode (4), we do not find a statistical correlation between $L_{\gamma}$ and $\Gamma_\gamma$ with a correlation coefficient of $r>0.5$ in a chance probability of $p<10^{-4}$ for other episodes. For the episode (4), the Pearson correlation analysis yields $r=0.534$ and $p=3.27\times10^{-4}$, indicating a tentative correlation between $L_{\gamma}$ and $\Gamma_\gamma$. Note that the above analysis is based on the data in time-bins of 10-day. The source experienced some extremely outbursts with timescale smaller than 10 days. Therefore, we also re-analyze the outbursts that have a peak luminosity $L_{\gamma}>2.6\times10^{47}$ erg s$^{-1}$ in Figure \ref{LC_all}(a) by using a time-bin of 1-day. Three outbursts (MJD [55381,55402], MJD [56083,56089], and MJD [56283,56297]) are included in our analysis, and their 1-day binned lightcurves and evolution of $\Gamma_\gamma$ as well as $\Gamma_{\gamma}$ as a function of $L_\gamma$ are presented in Figure \ref{flares}. A tentative correlation between $L_{\gamma}$ and $\Gamma_\gamma$ is found in flare (a) with $r=0.79$ and $p\sim 0.001$, and a trend of anticorrelation is presented in flare (b) with $r=-0.92$ and $p\sim0.01$, but no dependence of $\Gamma_\gamma$ on $L_{\gamma}$ is observed in flare (c). Our above results suggest that the spectral variation is not correlated with the $\gamma$-ray flux in the GeV band, similar to some flaring blazars (Nalewajko 2013).

\section{Searching for Possible QPO Signatures}

\begin{figure*}
\includegraphics[angle=0,scale=0.3]{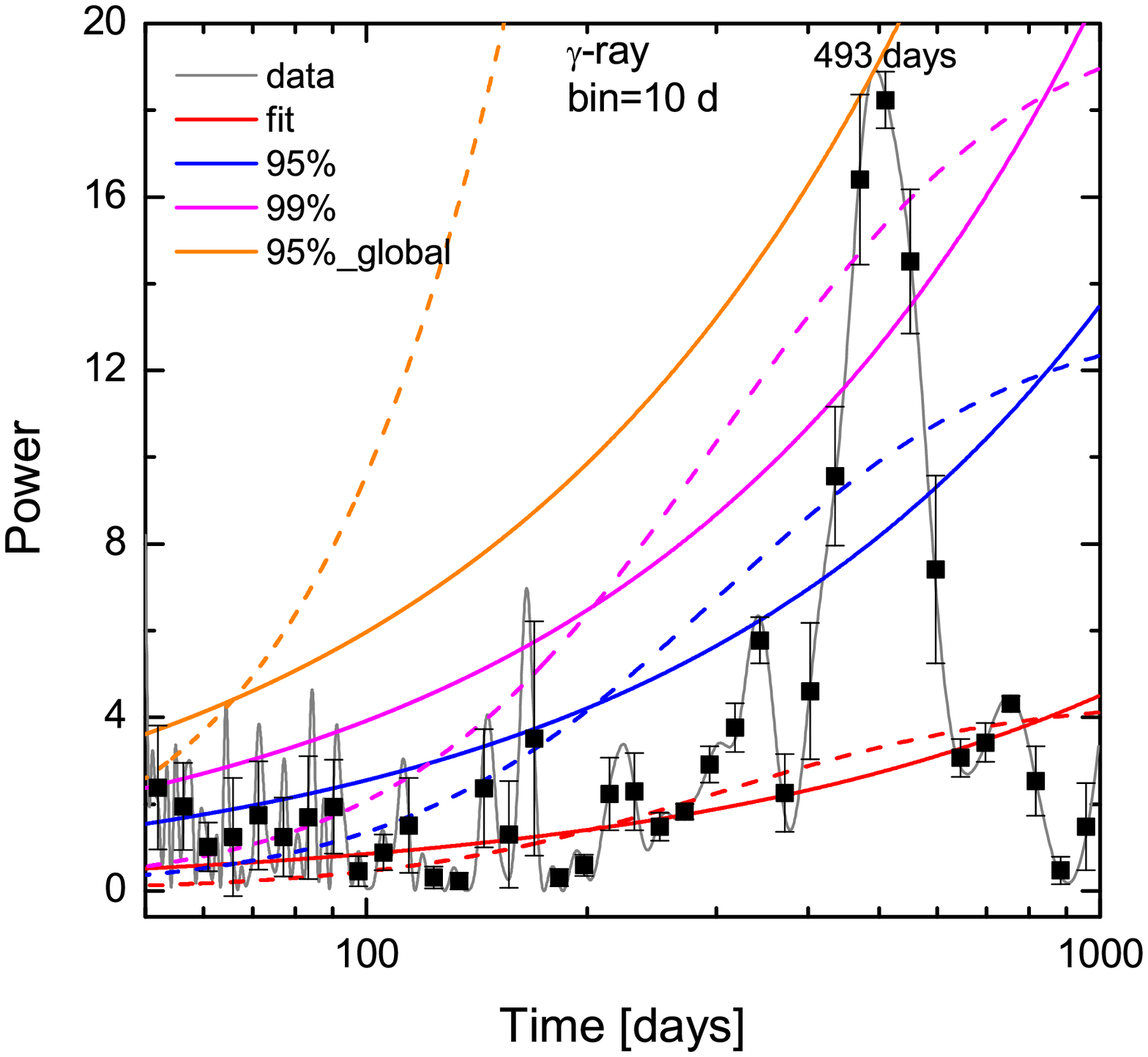}
\includegraphics[angle=0,scale=0.3]{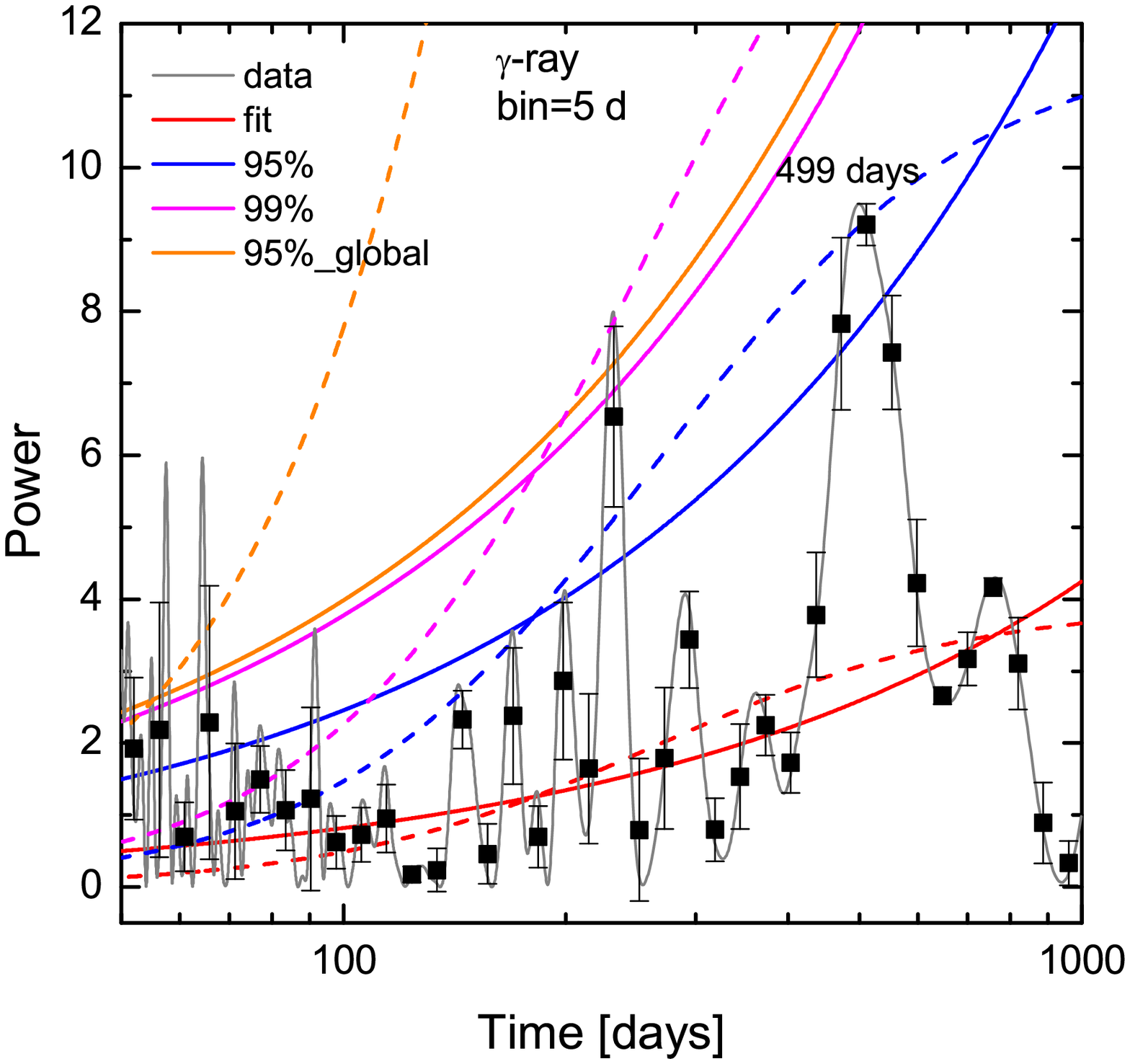}
\includegraphics[angle=0,scale=0.3]{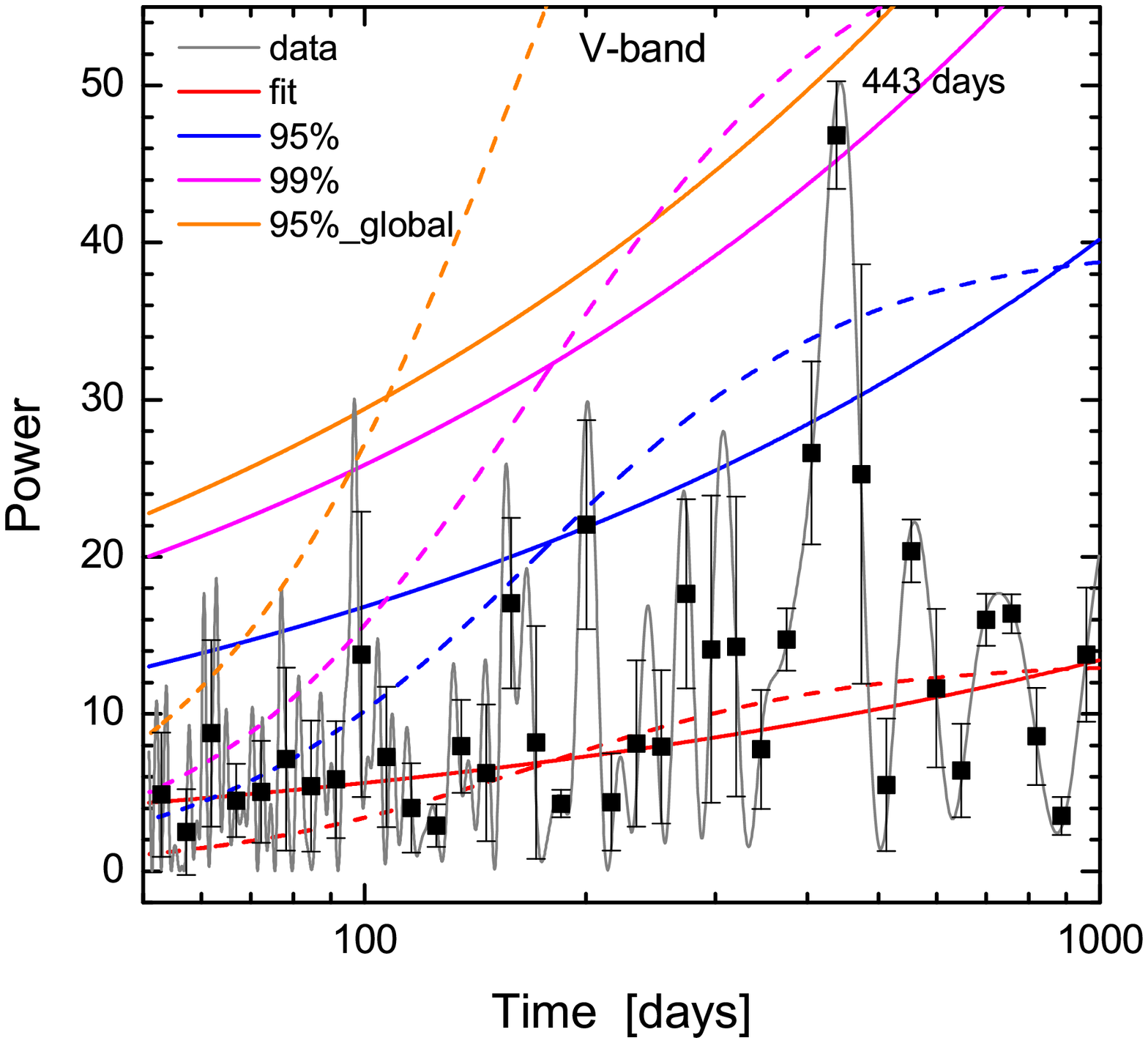}\\

\caption{LSPs (\emph{gray lines}) of lightcurves in $\gamma$-ray and optical bands. The \emph{black solid squares} indicate the re-binned data of LSPs. The best-fit noise spectrum are given in \emph{red lines}. Single frequency 95\% and 99\% confidence level lines are reported by \emph{blue and magenta lines}, and the global 95\% false-alarm levels of PL model are shown as \emph{orange lines}. \emph{Solid and dashed lines} indicate PL and AR1 models, respectively. }\label{LSP_correct}
\end{figure*}

To reveal the flux variation feature we analyze the power density spectrum (PDS) of the $\gamma$-ray, V-band, and radio lightcurves with the Lomb-Scargle Periodogram (LSP) algorithm (Lomb 1976; Scarle 1982). The LSP code is taken from Press et al. (1986). As suggested in Sandrinelli et al. (2017), we also use a power law (PL) and an auto-regression function of the first order (AR1) to fit the noise, respectively. The two models have been widely considered in the literature (e.g., K\"{o}nig \& Timmer 1997; Kelly et al. 2009; Edelson et al. 2013). The procedure of searching for possible periodic signals using Bayesian statistics described in Vaughan (2005, 2010) and Sandrinelli et al. (2017) is used to fit the PDS. Considering the large fluctuations in PDS, the PDS is re-binned in log scale, as presented in Figure \ref{LSP_correct}. The re-binned PDS is sued to search for possible periodic signals. We only consider the 50--1000 days interval during model fitting in order to avoid the very noisy high-frequency part of the PDS and the limit of the lightcurve length. Our analysis results are shown in Figure \ref{LSP_correct}. The following discussion is on the basis of the PL model fitting.

Interestingly, the LSPs of the $\gamma$-ray lightcurves show a peak at $493^{+71}_{-46}$ days in a 99\% confidence level and $499^{+40}_{-32}$ days in a 95\% confidence level for time-bins of 10-day and 5-day, respectively, where the errors are derived with the confidence level lines. Similar QPO signature is also observed in the LSP of the V-band lightcurve, i.e., $443^{+12}_{-13}$ days in a 99\% confidence level. As illustrated in Figure \ref{LSP_correct}, the AR1 model fitting yields the similar results to the PL model. In the radio 15 GHz, no similar QPO signature to $\gamma$-ray and optical bands is observed in a 95\% confidence level with both models.

We also evaluate the global significance of any peak in the PDS (see Vaughan 2010 and Sandrinelli et al. 2017 for a deeper discussion), and the global 95\% false-alarm levels are also presented in Figure \ref{LSP_correct}. In this case, the QPO signature at $\sim$490 days in the $\gamma$-ray band and $\sim$440 days in the V-band would not be singled out in a 95\% confidence level. However, as suggested in Sandrinelli et al. (2017), the modest significance peak at approximately the same frequency in different energy bands is still the interesting feature of sources. And the analysis results of DCF between the $\gamma$-ray flux with the optical and radio fluxes described in Section 3 also strengthen the possible QPO signatures in $\gamma$-ray and optical bands.

\section{Discussion}
The QPO signature is physically interesting. As mention in Section 1, similar QPO signal is found in some blazars, and its physical mechanisms are under debating. A super-massive black-hole binary (SMBHB) system may lead to the QPO signal since its Keplerian binary orbital motion would induce the periodic accretion perturbations (e.g., Sillanpaa et al. 1988; Lehto \& Valtonen 1996; Graham et al. 2015). The periodic accretion perturbations of disk should also result in the periodic oscillations of jet radiation. In this scenario, the periodicity of the system can be estimated with the Kepler's law, $P^2=4\pi a^3/G(m+M)$, where $a$ is the separation of the the two black holes, $M$ and $m$ are the masses of the two black holes, and $G$ is the Gravitational constant. The intrinsic orbital period of PMN J0948+0022 is given by $490/(1+z)=309$ days. The SMBH total mass in PMN J0948+0022, estimated with the Mg $_{\rm II}$ $\lambda$2798 and monochrome luminosity at 3000 ${\rm \AA}$ (Zhou et al. 2003), is $\sim 8.1\times10^8 M_{\odot}$. Assuming $(m+M)\sim 8.1\times10^8 M_{\odot}$, we have $a=1.24\times10^{16}$ cm ($\sim0.004$ pc), which is similar to the derived distance in PKS 1510--089 (Xie et al. 2002) and PG 1553+113 ($\sim0.005$ pc, Ackermann et al. 2015). Such a system would be sufficiently bound and its orbit would be shrunk by gravitational radiation. The merger timescale can be estimated with $t_m\sim 3\times 10^5 (m/M){M_8}^{-3}{a_{16}}^4$ years, where notation $Q_n=Q/10^n$ is used in the cgs units (Begelman et al. 1980). Thus, the merger timescale of the system would be $\sim 1.33\times 10^{3}(m/M)^{-1}$ years. Such kind of systems should be the potential objects for future gravitational wave detectors.

\begin{figure*}
\includegraphics[angle=0,scale=1.5]{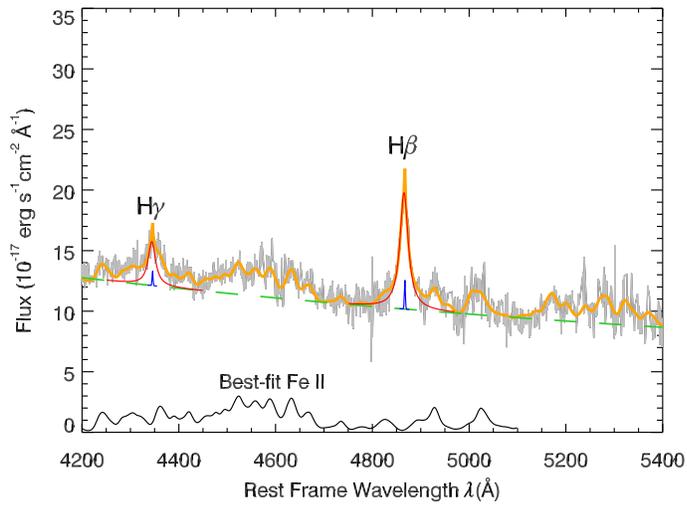}\\

\caption{The SDSS optical spectrum (\emph{grey line}) overplotted with the best-fit model (\emph{orange line}) near the H$\beta$ line range. The H$\beta$ and H$\gamma$ lines are fitted by a single Lorentzian profile as their broad components (\emph{red lines}) and a single Gaussian profile as their narrow components (\emph{blue lines}), respectively. The \emph{green dashed line} and \emph{black solid line} represent the best-fit power-law continuum and Fe{\sc\,ii} multiplets, where the Fe{\sc\,ii} multiplets are modeled using the templates in V\'{e}ron-Cetty et al. (2004). }\label{emission-lines}
\end{figure*}

In a SMBHB system with closer distance, the profile of the single-peaked spectral lines would be asymmetric since the tightly-bound black holes dynamically affect the broad-line region clouds as a single complex entity (Graham et al. 2015). We check wether the broad lines of PMN~J0948+0022 show a similar feature. PMN~J0948+0022 was spectroscopically observed in Sloan Digital Sky Survey (SDSS). After correcting for the Galactic extinction and transforming into the source rest frame, we reanalyze its SDSS spectrum following the same approach adopted in Yao et al. (2015). As shown in Figure \ref{emission-lines}, the asymmetric profiles of emission lines are not seen in PMN J0948+0022. This may give rise to an issue for explaining the QPO as a signature of a SMBHB system.

Alternative models were also proposed to interpret the QPO signal in blazars, such as the precession of jets (e.g., Stirling et al. 2003; Caproni et al. 2013) and the helical structure of jets (e.g., Conway \& Murphy 1993; Villata \& Raiteri 1999; Nakamura \& Meier 2004; Mohan \& Mangalam 2015). In a SMBHB system, misalignment between the accretion disk and the orbital plane of the secondary black hole could produce the torques that induces the precession of its jets. This mechanism is also used to explain the parsec-scale jet precession observed with the very-long-baseline interferometry (VLBI) technique (e.g., Caproni et al. 2013). In a single BH system, jet precession would be also produced by the misalignment of the rotation axes between accretion disk and Kerr black hole (Caproni et al. 2004). Similarly, helical jets or helical structures in jets can also be ascribed to a SMBHB system (e.g., Villata \& Raiteri 1999), or the hydrodynamical instabilities in magnetized jets (Hardee \& Rosen 1999). Being due to the variation of viewing angle to the jet axis, a QPO signature may be found for radiations from precessing jets or helical jets. We should point out that if the viewing angle effect, i.e., the change of Doppler boosting factor, is responsible for the QPO signature, the observed spectrum would harder when the source is brighter (Liu et al. 2010). However, we do not find such a feature in the GeV band.

So far there are several blazars that were suggested to have a year-like QPO (e.g., Sandrinelli et al. 2014, 2016; Ackermann et al. 2015; Zhang et al. 2017a, 2017b), however, the relative abundance of detected QPOs in quasars is lower (Charisi et al. 2016; Graham et al. 2015). Adding this GeV NLS1, it seems that the QPO signature is more common in jet-dominant sources since the oscillation would be amplified by the relativistic effects (Sandrinelli et al. 2017). In this respect, the QPOs in jet-dominant sources may more potentially be due to the jet instabilities or jet structures.

As discussed above, the mechanism that makes the QPO in PMN J0948+0022 is uncertain with the current observational data. Hence, the long-term monitoring observations, especially in multiwavelength, which prove or disprove the periodicity (see also Sandrinelli et al. 2017), or the detection of gravitational wave would be a robust probe for clarify these models.

\section{Conclusions}
We dealt with and analyzed the 8-year observation data of \emph{Fermi}/LAT and all the observation data of \emph{Swift}/XRT for the first confirmed GeV-NLS1 PMN J0948+0022, and also collected its long-term observation data in optical V-band and radio 15 GHz from the literature. It was found that this source shows significant flux variations in multi-wavelength. It demonstrates the similar characteristics to the typical FSRQs, but does not show a dependence of the spectral index variation to the flux variation in the GeV band. A QPO of $\sim 490$ days at a 99\% confidence level in the $\gamma$-ray lightcurve is found for PMN J0948+0022. Similar QPO signature is also observed in the optical V-band, i.e., $\sim 440$ days at a 99\% confidence level. The observed correlation between $\gamma$-ray and optical fluxes further strengthens the QPO behavior in PMN J0948+0022. No similar QPO signature is observed in the 15 GHz lightcurve, and no statistical correlation of variability between $\gamma$-ray and 15 GHz is found. We discussed the possible mechanisms that may produce the QPO. A tightly-bound SMBMB system may lead to the QPO, but we did not find any asymmetric profile feature of emission lines in PMN J0948+0022 as expected from such a system. The observations of a year-like QPO in jet-dominant sources may more potentially be due to the jet instabilities or jet structures. The long-term monitoring observations in multiwavelength, or gravitational wave observations in the future may clarify these plausible models.

\acknowledgments

We thank the anonymous referee for his/her valuable suggestions. We thank helpful discussion with Shuang-Nan Zhang and Yuan Liu. This research has made use of data from the OVRO 40-m monitoring program (Richards, J. L. et al. 2011, ApJS, 194, 29) which is supported in part by NASA grants NNX08AW31G, NNX11A043G, and NNX14AQ89G and NSF grants AST-0808050 and AST-1109911. This work is supported by the National Natural Science Foundation of China (grants 11573034, 11533003, 11363002, 11373036, 11463001), the National Basic Research Program (973 Programme) of China (grant 2014CB845800), and the Guangxi Science Foundation (2013GXNSFFA019001, 2014GXNSFAA118011). S. Yao. is supported by a KIAA-CAS Fellowship and Boya-PKU Fellowship.

\clearpage

\clearpage

\clearpage

\clearpage

\begin{thebibliography}

\bibitem[Abdo et al.(2009)]{2009ApJ...699..976A} Abdo, A.~A., Ackermann, M., Ajello, M., et al.\ 2009a, \apj, 699, 976


\bibitem[Abdo et al.(2009)]{2009ApJ...707L.142A} Abdo, A.~A., Ackermann, M., Ajello, M., et al.\ 2009b, \apjl, 707, L142


\bibitem[Abdo et al.(2009)]{2009ApJ...707..727A} Abdo, A.~A., Ackermann, M., Ajello, M., et al.\ 2009c, \apj, 707, 727


\bibitem[Ackermann et al.(2015)]{2015ApJ...813L..41A} Ackermann, M., Ajello, M., Albert, A., et al.\ 2015, \apjl, 813, L41


\bibitem[Arlen et al.(2013)]{2013ApJ...762...92A} Arlen, T., Aune, T., Beilicke, M., et al.\ 2013, \apj, 762, 92

\bibitem[Begelman et al.(1980)]{1980Natur.287..307B} Begelman, M.~C., Blandford, R.~D., \& Rees, M.~J.\ 1980, \nat, 287, 307

\bibitem[Caproni et al.(2013)]{2013MNRAS.428..280C} Caproni, A., Abraham, Z., \& Monteiro, H.\ 2013, \mnras, 428, 280


\bibitem[Caproni et al.(2004)]{2004ApJ...616L..99C} Caproni, A., Mosquera Cuesta, H.~J., \& Abraham, Z.\ 2004, \apjl, 616, L99

\bibitem[Charisi et al.(2016)]{2016MNRAS.463.2145C} Charisi, M., Bartos, I., Haiman, Z., et al.\ 2016, \mnras, 463, 2145

\bibitem[Conway \& Murphy(1993)]{1993ApJ...411...89C} Conway, J.~E., \& Murphy, D.~W.\ 1993, \apj, 411, 89


\bibitem[Cui(2004)]{2004ApJ...605..662C} Cui, W.\ 2004, \apj, 605, 662


\bibitem[D'Ammando et al.(2012)]{2012MNRAS.426..317D} D'Ammando, F., Orienti, M., Finke, J., et al.\ 2012, \mnras, 426, 317


\bibitem[D'Ammando et al.(2015)]{2015MNRAS.452..520D} D'Ammando, F., Orienti, M., Larsson, J., \& Giroletti, M.\ 2015, \mnras, 452, 520


\bibitem[Drake et al.(2009)]{2009ApJ...696..870D} Drake, A.~J., Djorgovski, S.~G., Mahabal, A., et al.\ 2009, \apj, 696, 870

\bibitem[Edelson \& Krolik(1988)]{1988ApJ...333..646E} Edelson, R.~A., \& Krolik, J.~H.\ 1988, \apj, 333, 646


\bibitem[Edelson et al.(2013)]{2013ApJ...766...16E} Edelson, R., Mushotzky, R., Vaughan, S., et al.\ 2013, \apj, 766, 16

\bibitem[Foschini et al.(2012)]{2012A&A...548A.106F} Foschini, L., Angelakis, E., Fuhrmann, L., et al.\ 2012, \aap, 548, A106

\bibitem[Ghisellini \& Tavecchio(2009)]{2009MNRAS.397..985G} Ghisellini, G., \& Tavecchio, F.\ 2009, \mnras, 397, 985


\bibitem[Graham et al.(2015)]{2015Natur.518...74G} Graham, M.~J., Djorgovski, S.~G., Stern, D., et al.\ 2015, \nat, 518, 74


\bibitem[Hardee \& Rosen(1999)]{1999ApJ...524..650H} Hardee, P.~E., \& Rosen, A.\ 1999, \apj, 524, 650

\bibitem[Jorstad et al.(2013)]{2013ApJ...773..147J} Jorstad, S.~G., Marscher, A.~P., Smith, P.~S., et al.\ 2013, \apj, 773, 147

\bibitem[Kalberla et al.(2005)]{2005A&A...440..775K} Kalberla, P.~M.~W., Burton, W.~B., Hartmann, D., et al.\ 2005, \aap, 440, 775

\bibitem[Kelly et al.(2009)]{2009ApJ...698..895K} Kelly, B.~C., Bechtold, J., \& Siemiginowska, A.\ 2009, \apj, 698, 895-910

\bibitem[Konig \& Timmer(1997)]{1997A&AS..124..589K} Konig, M., \& Timmer, J.\ 1997, \aaps, 124,

\bibitem[Lehto \& Valtonen(1996)]{1996ApJ...460..207L} Lehto, H.~J., \& Valtonen, M.~J.\ 1996, \apj, 460, 207


\bibitem[Liu et al.(2010)]{2010A&A...516A..16L} Liu, T., Liang, E.-W., Gu, W.-M., et al.\ 2010, \aap, 516, A16

\bibitem[Lomb(1976)]{1976Ap&SS..39..447L} Lomb, N.~R.\ 1976, \apss, 39, 447


\bibitem[Mahabal et al.(2011)]{2011BASI...39..387M} Mahabal, A.~A., Djorgovski, S.~G., Drake, A.~J., et al.\ 2011, Bulletin of the Astronomical Society of India, 39, 387


\bibitem[Massaro et al.(2008)]{2008A&A...478..395M} Massaro, F., Tramacere, A., Cavaliere, A., Perri, M., \& Giommi, P.\ 2008, \aap, 478, 395


\bibitem[Mohan \& Mangalam(2015)]{2015ApJ...805...91M} Mohan, P., \& Mangalam, A.\ 2015, \apj, 805, 91


\bibitem[Nakamura \& Meier(2004)]{2004ApJ...617..123N} Nakamura, M., \& Meier, D.~L.\ 2004, \apj, 617, 123


\bibitem[Nalewajko(2013)]{2013MNRAS.430.1324N} Nalewajko, K.\ 2013, \mnras, 430, 1324


\bibitem[Paliya et al.(2015)]{2015AJ....149...41P} Paliya, V.~S., Stalin, C.~S., \& Ravikumar, C.~D.\ 2015, \aj, 149, 41

\bibitem[Press et al.(1986)]{} Press, W.~H., Teukolsky, S.~A., Vetterling, W.~T., \& Flannery, B,~P.\ 1986, Numerical Recipes in Fortran 77


\bibitem[Richards et al.(2011)]{2011ApJS..194...29R} Richards, J.~L., Max-Moerbeck, W., Pavlidou, V., et al.\ 2011, \apjs, 194, 29


\bibitem[Sandrinelli et al.(2016)]{2016AJ....151...54S} Sandrinelli, A., Covino, S., Dotti, M., \& Treves, A.\ 2016, \aj, 151, 54


\bibitem[Sandrinelli et al.(2014)]{2014ApJ...793L...1S} Sandrinelli, A., Covino, S., \& Treves, A.\ 2014, \apjl, 793, L1


\bibitem[Sandrinelli et al.(2017)]{2017A&A...600A.132S} Sandrinelli, A., Covino, S., Treves, A., et al.\ 2017, \aap, 600, A132

\bibitem[Scargle(1982)]{1982ApJ...263..835S} Scargle, J.~D.\ 1982, \apj, 263, 835


\bibitem[Sillanpaa et al.(1988)]{1988ApJ...325..628S} Sillanpaa, A., Haarala, S., Valtonen, M.~J., Sundelius, B., \& Byrd, G.~G.\ 1988, \apj, 325, 628


\bibitem[Sillanpaa et al.(1996)]{1996A&A...305L..17S} Sillanpaa, A., Takalo, L.~O., Pursimo, T., et al.\ 1996, \aap, 305, L17


\bibitem[Stirling et al.(2003)]{2003MNRAS.341..405S} Stirling, A.~M., Cawthorne, T.~V., Stevens, J.~A., et al.\ 2003, \mnras, 341, 405


\bibitem[Sun et al.(2014)]{2014JApA...35..457S} Sun, X.~N., Zhang, J., Lu, Y., Liang, E.~W., \& Zhang, S.~N.\ 2014, Journal of Astrophysics and Astronomy, 35, 457


\bibitem[Sun et al.(2015)]{2015ApJ...798...43S} Sun, X.-N., Zhang, J., Lin, D.-B., et al.\ 2015, \apj, 798, 43


\bibitem[Tramacere et al.(2009)]{2009A&A...501..879T} Tramacere, A., Giommi, P., Perri, M., Verrecchia, F., \& Tosti, G.\ 2009, \aap, 501, 879


\bibitem[Vaughan(2010)]{2010MNRAS.402..307V} Vaughan, S.\ 2010, \mnras, 402, 307


\bibitem[Vaughan(2005)]{2005A&A...431..391V} Vaughan, S.\ 2005, \aap, 431, 391

\bibitem[V{\'e}ron-Cetty et al.(2004)]{2004A&A...417..515V} V{\'e}ron-Cetty, M.-P., Joly, M., \& V{\'e}ron, P.\ 2004, \aap, 417, 515


\bibitem[Villata \& Raiteri(1999)]{1999A&A...347...30V} Villata, M., \& Raiteri, C.~M.\ 1999, \aap, 347, 30


\bibitem[Xie et al.(2002)]{2002MNRAS.334..459X} Xie, G.~Z., Liang, E.~W., Zhou, S.~B., et al.\ 2002, \mnras, 334, 459

\bibitem[Yao et al.(2015)]{2015MNRAS.454L..16Y} Yao, S., Yuan, W., Zhou, H., et al.\ 2015, \mnras, 454, L16

\bibitem[Yuan et al.(2008)]{2008ApJ...685..801Y} Yuan, W., Zhou, H.~Y., Komossa, S., et al.\ 2008, \apj, 685, 801-827


\bibitem[Zhang et al.(2013)]{2013IAUS..290..359Z} Zhang, J., Sun, X.~N., Zhang, S.~N., \& Liang, E.~W.\ 2013, Feeding Compact Objects: Accretion on All Scales, 290, 359


\bibitem[Zhang et al.(2017)]{2017ApJ...835..260Z} Zhang, P.-f., Yan, D.-h., Liao, N.-h., \& Wang, J.-c.\ 2017a, \apj, 835, 260

\bibitem[Zhang et al.(2017)]{2017ApJ...842...10Z} Zhang, P.-f., Yan, D.-h., Liao, N.-h., et al.\ 2017b, \apj, 842, 10

\bibitem[Zhou et al.(2003)]{2003ApJ...584..147Z} Zhou, H.-Y., Wang, T.-G., Dong, X.-B., Zhou, Y.-Y., \& Li, C.\ 2003, \apj, 584, 147




\end{thebibliography}
\end{document}